\documentclass[twocolumn]{aastex631}

\usepackage[T1]{fontenc}
\usepackage{ae,aecompl}
\usepackage{natbib}

\usepackage{graphicx}	
\usepackage{amsmath}	
\usepackage{amssymb}	
\usepackage{bm}		
\usepackage{booktabs}
\usepackage{mathtools}
\usepackage{makecell}
\usepackage{bm}
\usepackage{float}
\usepackage[utf8x]{inputenc} 
\usepackage{color,soul}
\usepackage{breqn}

\newcommand{\fesc}{\ensuremath{f_{\rm esc}}}

\newcommand{\wlya}{\ensuremath{W_{\lambda}}(\rm Ly\ensuremath{\alpha})}
\newcommand{\woiii}{\ensuremath{W_{\lambda}}(\rm [O~\textsc{iii}])}

\newcommand{\luv}{\ensuremath{L_{\rm UV}}}

\newcommand{\zspec}{\ensuremath{z_{\rm spec}}}

\newcommand{\mstar}{\ensuremath{\rm M_{\rm *}}}
\newcommand{\muv}{\ensuremath{\rm M_{\rm UV}}}
\newcommand{\msun}{\ensuremath{M_{\odot}}}
\newcommand{\xiion}{\ensuremath{\xi_{\rm ion}}}
\newcommand{\xiiono}{\ensuremath{\xi_{\rm ion,0}}}
\newcommand{\oh}{\ensuremath{12+\textrm{log(O/H)}}}
\newcommand{\av}{\ensuremath{\textrm{A}_{\rm V}}}
\newcommand{\lxiion}{\textrm{log}\ensuremath{_{10}(\xi_{\rm ion}/\textrm{erg\:Hz}^{-1})}}
\newcommand{\lxiiono}{\textrm{log}\ensuremath{_{10}(\xi_{\rm ion,0}/\textrm{erg\:Hz}^{-1})}}

\newcommand{\lya}{Ly$\alpha$}
\newcommand{\oiii}{[O~\textsc{iii}]\ensuremath{\lambda\lambda4959,5007}}
\newcommand{\oiiif}{[O~\textsc{iii}]\ensuremath{\lambda5007}}
\newcommand{\oii}{[O~\textsc{ii}]\ensuremath{\lambda\lambda3726,3729}}
\newcommand{\hb}{H\ensuremath{\beta}}
\newcommand{\ha}{H\ensuremath{\alpha}}
\newcommand{\neiii}{[Ne~\textsc{iii}]\ensuremath{\lambda3870}}
\newcommand{\nii}{[N~\textsc{ii}]\ensuremath{\lambda6585}}

\shorttitle{\xiion{} with JWST/NIRSpec}
\shortauthors{Pahl et al.}

\begin{document}
	
	\title[.]{A spectroscopic analysis of the ionizing photon production efficiency in JADES and CEERS: implications for the ionizing photon budget}
	
	\author[0000-0003-4464-4505]{Anthony J. Pahl}
	\altaffiliation{Carnegie Fellow}
	\affiliation{The Observatories of the Carnegie Institution for Science, 813 Santa Barbara Street, Pasadena, CA 91101, USA}
	
	\author[0000-0001-8426-1141]{Michael W. Topping}
	\affiliation{Steward Observatory, University of Arizona, 933 N. Cherry Avenue, Tucson, AZ 85721, USA}
	
	\author[0000-0003-3509-4855]{Alice Shapley}
	\affiliation{Department of Physics and Astronomy, University of California, Los Angeles, CA 90095, USA}
	
	\author[0000-0003-4792-9119]{Ryan Sanders}
	\affiliation{University of Kentucky, 506 Library Drive, Lexington, KY, 40506, USA}
	
	\author[0000-0001-9687-4973]{Naveen A. Reddy}
	\affiliation{Department of Physics and Astronomy, University of California Riverside, Riverside, CA 92521, USA}
	
	\author[0000-0003-1249-6392]{Leonardo Clarke}
	\affiliation{Department of Physics and Astronomy, University of California, Los Angeles, CA 90095, USA}
	
	\author{Emily Kehoe}
	\affiliation{Department of Physics and Astronomy, University of California, Los Angeles, CA 90095, USA}
	
	\author{Trinity Bento}
	\affiliation{Department of Physics and Astronomy, University of California, Los Angeles, CA 90095, USA}
	
	\author[0000-0003-2680-005X]{Gabe Brammer}
	\affiliation{Cosmic Dawn Center (DAWN), Denmark}
	\affiliation{Niels Bohr Institute, University of Copenhagen, Jagtvej 128, DK-2200 Copenhagen N, Denmark}
	
	\begin{abstract}
		We have used a combined sample of JADES and CEERS objects in order to constrain ionizing photon production efficiency (\xiion) from \textit{JWST}/NIRSpec and \textit{JWST}/NIRCam data. We examine 163 objects at $1.06<z<6.71$ with significant ($3\sigma$) spectroscopic detections of \ha{} and \hb{} in order to constrain intrinsic \ha{} luminosities corrected from nebular dust attenuation via Balmer decrements. We constrain dust-corrected UV luminosities from best-fit spectral-energy distribution modeling. We find a sample median $\lxiiono=25.29^{+0.29}_{-0.37}$, assuming \fesc=0 for the escape fraction of Lyman continuum emission. We find significant correlation between \xiiono{} and $z$, with objects at $z>4.64$ having median $\lxiiono=25.38^{+0.38}_{-0.38}$, with those below having $\lxiiono=25.24^{+0.30}_{-0.33}$. We also find significant, positive correlations between \xiiono{} and \luv{}; \woiii{}; \oiiif{}/\oii{}; and inverse correlations with metallicity. In contrast with some previous results, we find no trends between \xiiono{} and stellar mass, stellar dust attenuation, or UV slope. Applying a multivariate fit to \xiiono{}, $z$, and \muv{} to an empirically-motivated model of reionization, and folding in \fesc{} estimates from direct observations of the Lyman continuum at $z\sim3$ from the Keck Lyman Continuum Spectroscopic survey, we find that the number of ionizing photons entering the IGM causes reionization to end at $z\sim5-7$.
	\end{abstract}
	
	\keywords{Galaxy evolution (594), High-redshift galaxies (734), Reionization (1383), Near infrared astronomy (1093)}

	\section{Introduction} \label{sec:intro}
	
	The epoch of reionization, categorized by the Hydrogen in the intergalactic medium (IGM) transitioning from neutral to ionized, has been thought to conclude at $z\sim5.5-6.0$ based on the emergent UV spectra of distant quasars \citep{fanObservationalConstraintsCosmic2006,beckerMeanFreePath2021,bosmanHydrogenReionizationEnds2022}. Considering that the number density of quasars appreciates significantly only after this purported endpoint in cosmic time \citep[e.g.,][]{shenBolometricQuasarLuminosity2020}, young, massive stars from star-forming galaxies are thought to be the driving forces behind the process.
	
	Considerable uncertainty remains in the wholesale ionizing photon emissivity injected into the IGM from star-forming galaxies as a function of cosmic time, despite the revolutionary effects that the \textit{JWST} has had on our understanding of early galaxies. This emissivity can be parameterized  as a function of three parameters: the cosmic star-formation rate density $\rho_{\rm SFR}$, the ionizing photon production efficiency \xiion{}, and the escape fraction \fesc{} of Lyman continuum (LyC) photons from the galaxies in which they were produced \citep{robertsonCosmicReionizationEarly2015}. \textit{JWST} has been instrumental in mapping the UVLF well into the reionization epoch at $z=8-13$ \citep[e.g.,][]{donnanEvolutionGalaxyUV2023,harikaneComprehensiveStudyGalaxies2023,finkelsteinCompleteCEERSEarly2023,robertsonEarliestGalaxiesJADES2023,leungNGDEEPEpochFaint2023}, reducing past uncertainties on $\rho_{\rm SFR}$, but the faint-end slope remains particularly uncertain at $z>10$. 
	
	\textit{JWST} is also enabling new estimates of \xiion{} with spectroscopic and photometric coverage of nebular emission lines, and new surveys are finding elevated \xiion{} than what has been previously assumed in the reionization epoch \citep[e.g., $\lxiion=25.2$,][]{robertsonCosmicReionizationEarly2015}. Analyses are finding anywhere from $\lxiion\sim25.3$ \citep{mattheeEIGERIIFirst2023,endsleyStarformingIonizingProperties2023,prieto-lyonProductionIonizingPhotons2023} to $\lxiion\sim25.5$ and above \citep{jungCEERSDiversityLymanAlpha2023,simmondsIonizingPhotonProduction2023,simmondsLowmassBurstyGalaxies2024,atekMostPhotonsThat2024} depending on measurement methodology and sample selection function. Two large-scale analyses of the \textit{JWST} Advanced Deep Extragalactic Survey \citep[JADES,][]{eisensteinOverviewJWSTAdvanced2023}, examining the fluxes of \ha{} and \oiii{} emission lines via photometric excesses and spectral-energy density (SED) fitting, find opposite trends between \xiion{} and UV brightness: either brighter galaxies are more efficient producers of ionizing photons \citep{endsleyStarformingIonizingProperties2023}, or objects with fainter UV luminosities are more efficient \citep{simmondsLowmassBurstyGalaxies2024}.
	
	 The third term, \fesc{}, cannot be directly constrained at $z>4$ due to absorption of LyC radiation from trace amounts of neutral Hydrogen left in the IGM after reionization \citep{vanzellaDetectionIonizingRadiation2012}. Spectroscopic surveys of analogues to reionization-era galaxies have been performed in order to determine the key galaxy properties in inferring this parameter at higher redshifts.
	 At $z\sim0$, the IGM is almost entirely ionized, and allows for confident measurements of the LyC, and thus \fesc{}, of individual objects, but the galaxy conditions in the local Universe make finding individual leakers quite difficult. The Low-Redshift Lyman Continuum Survey (LzLCS)  achieved success by selecting blue, compact, star-forming galaxies with high [OIII]/[OII], and showed that \fesc{} at $z\sim0$ is correlated with UV slope ($\beta$), velocity separation of double-peaked \lya{}, and star-formation rate (SFR) surface density \citep{chisholmFarUltravioletContinuumSlope2022,fluryLowredshiftLymanContinuum2022,fluryLowredshiftLymanContinuum2022a}
	 The Keck Lyman Continuum Spectroscopic survey (KLCS) represents a signpost of the ISM and CGM conditions of Lyman break selected galaxies at $z\sim3$. At this epoch and for galaxies with $\luv\sim L^*_{\rm UV}$, \fesc{} is $0.06\pm0.01$ on average, and \fesc{} is a associated with both UV luminosity (\luv{}) and the equivalent width of \lya{} (\wlya{}), with weaker trends between \fesc{} and stellar mass (\mstar{}), E(B-V), and the velocity offset in the red peak of \lya{} emission \citep{steidelKeckLymanContinuum2018,pahlUncontaminatedMeasurementEscaping2021,pahlConnectionEscapeIonizing2023,pahlLyAlphaProfile2024}.
	 	 
	 Given the novel constraints on both the UVLF and \xiion{} that \textit{JWST} has enabled, \citet{munozReionizationJWSTPhoton2024} synthesized these first constraints and found a puzzling result: given the \xiion{} values inferred from SED fits to 677 JADES galaxies \citep{simmondsLowmassBurstyGalaxies2024} and the \fesc{}-$\beta$ relation from the LzLCS \citep{chisholmFarUltravioletContinuumSlope2022}, far too many ionizing photons are produced under reasonable assumptions, and reionization ends much earlier ($z\sim9$) than expected from observational constraints \citep[e.g.,][]{planckcollaborationPlanck2015Results2016,nakaneLyaEmission132024}. Modifying \fesc{} to a constant 20\% did not entirely relieve this tension. Thus, an ionizing-budget crisis emerges: there appear to be too many ionizing photons produced by faint galaxies directly observed with \textit{JWST}, assuming an appreciable fraction of them ($\gtrsim3$\%) escape into the IGM.
	 
	 Still, we lack large-scale, spectroscopic analyses of \xiion{} compiled from multiple early \textit{JWST} data releases. Detection of multiple Balmer lines is especially critical for properly correcting for nebular dust attenuation when estimating \xiion{}, which can performed via measurement of the Balmer decrement \citep[e.g.,][]{shivaeiMOSDEFSurveyDirect2018}.
	 In this work, we examine the \xiion{} assumptions that result in an ionizing-budget crisis by performing a comprehensive analysis of galaxies in two early \textit{JWST} programs: the Cosmic Evolution Early Release Science Survey \citep[CEERS,][]{finkelsteinCompleteCEERSEarly2023,arrabalharoSpectroscopicConfirmationCEERS2023} and JADES. Using the combined \textit{JWST}/NIRSpec sample first presented in \citet{clarkeStarFormingMainSequence2024}, we focus on objects with spectroscopic coverage and significant detections of both \ha{} and \hb{}. Using this comprehensive, spectroscopic sample with a simple selection function, we aim to accurately determine the evolution of \xiion{} both with $z$ and \muv{}, and perform a careful accounting of sample completeness in order to understand the limitations of the galaxies so-far observed spectroscopically by \textit{JWST}. We also examine the reionization models as presented by \citet{munozReionizationJWSTPhoton2024} with our novel constraints on \xiion{}, while introducing empirical \fesc{} assumptions drawn directly from high-redshift ($z\sim3$) observations of the KLCS.
	
	The paper is structured as follows. In Section \ref{sec:data}, we review the data and reduction techniques of the combined CEERS and JADES dataset. In Section \ref{sec:methods}, we the methodology for emission-line and SED fitting, and \xiion{} estimation, as well as measurements of $\beta$ and metallicity. In Section \ref{sec:res}, we present the \xiion{} measurements and examine the evolution of \xiion{} with a respect to a number of galaxy properties. We discuss comparisons to results from the literature in Section \ref{sec:discussion}, alongside implications for reionization within the framework of \citet{munozReionizationJWSTPhoton2024}. We conclude in Section \ref{sec:summary}.
	
	Throughout this paper, we adopt a standard $\Lambda$CDM cosmology with $\Omega_m$ = 0.3, $\Omega_{\Lambda}$ = 0.7 and $H_0$ = 70 $\textrm{km\,s}^{-1}\textrm{Mpc}^{-1}$. We also employ the AB magnitude system \citep{okeSecondaryStandardStars1983}. Additionally, we use the solar abundances reported by \citet{asplundChemicalCompositionSun2009} with $\oh=8.69$ and the solar metallicity as Z$_{\rm \odot}=0.014$.
	
	\section{Data and reduction} \label{sec:data}
	
	We utilize available \textit{JWST}/NIRSpec survey data that targets high ($z\sim1-6$) redshift	sources in order to constrain \xiion{} directly from observed Balmer emission lines and spectral energy distributions (SEDs) that are well sampled in the rest-UV. Here, we describe the data originally observed within the CEERS \citep[Program ID:1345]{finkelsteinCompleteCEERSEarly2023,arrabalharoSpectroscopicConfirmationCEERS2023} and JADES \citep[Program ID: 1210]{eisensteinOverviewJWSTAdvanced2023} programs and the reduction techniques, originally presented in \citet{sandersExcitationIonizationProperties2023} and \citet{clarkeStarFormingMainSequence2024}.
	
	\subsection{Spectroscopy}
	\subsubsection{CEERS}
	Our analysis draws from six CEERS NIRSpec Micro-Shutter Assembly (MSA) pointings in the AEGIS field, using the grating/filter combinations G140M/F100LP, G235M/F170LP, and G395M/F290LP. These settings provide a spectral resolution of $R\sim1000$ across a wavelength range of approximately $1-5 \mu$m. Each pointing was observed for 3107 seconds per grating/filter combination, with a 3-point nod pattern employed for each 3-shutter MSA slit. Altogether, the 6 pointings covered 321 slits and targeted 318 unique objects.
	
	The CEERS data were reduced using the STScI \textit{JWST} pipeline\footnote{\url{https://jwst-pipeline.readthedocs.io/en/latest/index.html}}. For further details on the 2D data reduction, refer to \citet{shapleyJWSTNIRSpecBalmerline2023} and \citet{sandersExcitationIonizationProperties2023}. Briefly, after applying the pipeline to the data, the exposures were combined to create 310 individual 2D spectra. These were then converted to 1D spectra as described in \citet{sandersExcitationIonizationProperties2023}, yielding 252 1D spectra. The resulting spectra were corrected for wavelength-dependent slit losses \citep[see][]{reddyPaschenlineConstraintsDust2023} and a flux calibration was performed such that the flux densities aligned with broadband photometry.
	
	
	
	\subsubsection{JADES}
	We utilized publicly available data from the JADES program within the GOODS-S extragalactic legacy field. The complete data release included NIRSpec spectra obtained using PRISM/CLEAR, G140M/F070LP, G235M/F170LP, G395M/F290LP, and G395H/F290LP grating/filter configurations, while in this analysis we focused on the observations using G140M/F070LP, G235M/F170LP, and G395M/F290LP. The NIRSpec observations spanned three different visits, and targeted 198 galaxies in the aforementioned grating/filters, 117 of which had recovered redshifts as presented in \citet{clarkeStarFormingMainSequence2024}. The objects were observed with one, two, or three visits, with 2.3\:hr exposures times per visit. A more detailed description of the JADES NIRSpec observations is provided by \citet{bunkerJADESNIRSpecInitial2023}.
	
	The reduced 2D JADES NIRSpec spectra were sourced from the DAWN JWST Archive (DJA)\footnote{\url{https://dawn-cph.github.io/dja/spectroscopy/nirspec/}}, employing a reduction process similar to that used for CEERS NIRSpec data. The reduction was conducted using a custom pipeline, \texttt{MsaExp}\footnote{\url{https://github.com/gbrammer/msaexp}}, which takes Stage 2 data products from the MAST database\footnote{\url{https://archive.stsci.edu/hlsp/jades}} as inputs and performs wavelength calibrations, flat-fielding, photometric calibrations, and point-source slit-loss corrections on each individual exposure. Each 2D exposure was then co-added and combined into a final 2D spectrum for each object. Further details on the reduction of the 2D spectra can be found in \citet{heintzExtremeDampedLyman2023} and \citet{heintzJWSTPRIMALLegacySurvey2024}. The 1D spectra were extracted in a similar way to that of the CEERS data, following the procedure outlined in \citet{clarkeStarFormingMainSequence2024} and consistent with that in \citet{sandersExcitationIonizationProperties2023}. Slit-loss corrections beyond the default point-source were not applied to the JADES spectra due to the uncertainty of the slit positioning between visits. A multiplicative scaling factor was applied to G140M and G395M gratings (leaving G235 fixed), ensuring that targets with the same exposure time should recover the same relative sensitivity between gratings. Finally, all the gratings were scaled by a median factor to match the photometry on average.

	\subsection{Photometry}
	
	For both the CEERS and JADES samples, we utilized publicly accessible multi-wavelength photometric catalogs curated by G. Brammer\footnote{\url{https://s3.amazonaws.com/grizli-v2/JwstMosaics/v7/index.html}}. The CEERS catalog features 7 HST bands (F435W, F606W, F814W, F105W, F125W, F140W, and F160W) and 7 JWST/NIRCam bands (F115W, F150W, F200W, F277W, F356W, F410M, and F444W). The JADES catalog includes 9 HST bands (F435W, F606W, F775W, F814W, F850LP, F105W, F125W, F140W, and F160W) and 14 JWST/NIRCam bands, which comprise broadband filters (F090W, F115W, F150W, F200W, F277W, F356W, and F444W) as well as medium-band filters (F182M, F210M, F335M, F410M, F430M, F460M, and F480M). These medium-band filters are part of the JEMS program \citep{williamsJEMSDeepMediumband2023}. We made use of SEDs cataloged by the 3D-HST team \citep{skelton3DHSTWFC3selectedPhotometric2014,momcheva3DHSTSURVEYHUBBLE2016} in the AEGIS and GOODS-S fields for CEERS and JADES objects, respectively, that did not have NIRCam coverage. These catalogs include ground-based and \textit{HST} optical and near-IR photometry, alongside Spitzer/IRAC measurements ranging from $3.6-8.0 \mu$m.
	
	Within the CEERS sample, 231 galaxies have measured, spectroscopic redshifts, with 217 having a combination of NIRCam and/or 3DHST measurements. Of the 117 JADES objects with spectroscopic redshifts, 113 have either NIRCam or 3DHST photometry available, with three having only \textit{HST} photometry.
		
	\section{Physical Properties and Measurements} \label{sec:methods}
	
	\subsection{Emission line fluxes and SED Fitting} \label{sec:lineflux}
	
	Emission line fluxes were measured via the method described in \citet{clarkeStarFormingMainSequence2024}. In summary, lines were fit using \texttt{scipy.optimize.curve\_fit} to perform a $\chi^2$ minimization using a single Gaussian profile, save for adjacent emission lines that were fit simultaneously (such as \ha{} and \nii{}). The 68\% confidence interval of the line fluxes were obtained via Monte Carlo simulations, perturbing the flux density spectrum according to the error spectrum 500 times and determining the 16th and 84th percentiles of the resulting distributions.
	
	The spectral energy distribution (SED) was fit using the full suite of available photometry, which included \textit{JWST}/NIRCam, \textit{HST}/WFC3, \textit{Spitzer}/IRAC, and ground-based surveys to cover the rest-UV to the rest-infrared. The SED fitting was described in \citet{shapleyJWSTNIRSpecBalmerline2023} and \citet{clarkeStarFormingMainSequence2024}, and used the FAST software described in \citet{kriekMOSFIREDEEPEVOLUTION2015}. Following \citet{reddyHDUVSurveyRevised2018} and \citet{duRedshiftEvolutionRestUV2018}, we performed the SED fits using a combination of a fixed metallicity of 1.4 $Z_{\rm \odot}$ and a \citet{calzettiDustContentOpacity2000} extinction curve, or a fixed metallicity of 0.27 $Z_{\rm \odot}$ and an SMC extinction curve \citep{gordonQuantitativeComparisonSmall2003}.  For galaxies at $z\leq1.4$, we assumed $1.4Z_{\rm \odot}$+Calzetti. For galaxies at $1.4<z\leq2.7$, we assumed 0.27$Z_{\rm \odot}$+SMC for galaxies with  log$(\mstar/\msun)<10.45$ and $1.4Z_{\rm \odot}$+Calzetti for galaxies with  log$(\mstar/\msun)\geq10.45$. For galaxies at $z>3.4$, we assumed 0.27$Z_{\rm \odot}$+SMC. We assumed a delayed-$\tau$ star-formation history and a \citet{chabrierGalacticStellarSubstellar2003} initial mass function.
	
	Photometric bands with coverage of strong nebular emission lines were corrected from flux excesses using the emission-line fluxes detected at the $>5\sigma$ level, using the method described in \citet{sandersMOSDEFSurveyEvolution2021}. The emission-line-corrected SEDs were then re-fit using the same modelling methodology, and these final best-fit SEDs were used to model the continuum in the final run of the emission line fitting.

	\subsection{Selection criteria} \label{sec:selection}
	
	In order to construct a sample of galaxies for exploring ionizing photon production efficiencies at high redshift, we required an estimate of the intrinsic (i.e., dust corrected) \ha{} emission luminosity and an estimate of the non-ionizing rest-UV continuum. We first restricted the redshift range of the sample to $0.67\leq \zspec \leq 6.71$, selecting all objects that have both \ha{} and \hb{} covered by \textit{JWST}/NIRSpec. Both \ha{} and \hb{} fluxes are necessary to perform corrections on \ha{} line luminosity for dust attenuation via the intrinsic \ha{}/\hb{} Balmer decrement, described in Section \ref{sec:int_ha}. To this end, we also required a $>3\sigma$ detection in both \ha{} and \hb{}. We leave analysis of higher redshift sources via simultaneous detections of \hb{} and H$\gamma$ to future work. We required both photometric coverage at $\lambda_{\rm rest}<2000$\AA{} in order to constrain the continuum near $\lambda_{\rm rest}\sim1500$\AA{} and a robust, multi-wavelength SED such that the stellar dust attenuation, needed to correct \luv{}, could be estimated from SED fitting. Finally, we remove candidate AGN from our sample on the basis of  [NII]~$\lambda6563/$\ha$>0.5$ or evidence of both a broad and narrow component within the \ha{} line-profile shape \citep{shapleyJWSTNIRSpecBalmerline2023}. Our final analysis sample contains 163 galaxies (118 in CEERS and 45 in JADES), and features objects in the range $1.06\leq z\leq6.61$. We assumed 0.27$Z_{\rm \odot}$+SMC for 153 of the galaxies, and $1.4Z_{\rm \odot}$+Calzetti for nine of the galaxies.
	
\begin{figure} %
		\centering
		\includegraphics[width=\columnwidth]{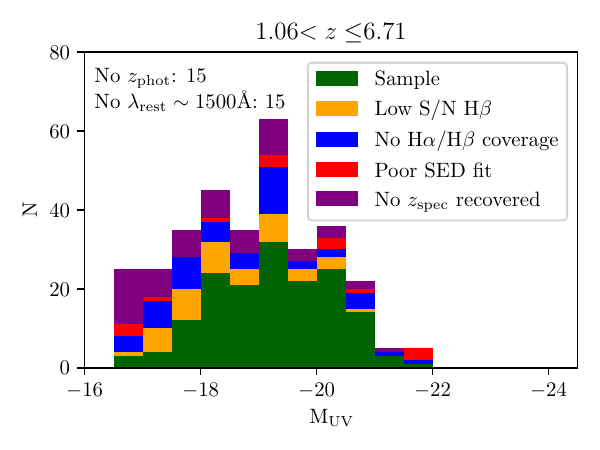}
		\caption{Sample completeness of the combined JADES/CEERS NIRSpec sample. The final selected sample (N=163) is shown in green. Objects targeted with JWST/NIRSpec were omitted from our analysis if \zspec{} was not recovered (purple, N=92), \hb{} S/N$<3$ (orange, N=48), the spectra did not cover \ha{} and/or \hb{} (blue, N=59), or the SED fit to the photometry was poor (red, N=18). Additionally, 15 objects had neither \zspec{} or $z_{\rm phot}$ measurements available, and 15 objects had no photometry near rest frame 1500\AA{}, and thus no constraints on \muv{}. Our analysis sample is largely complete (66\%) at $\muv<-19.5$, and falls to $\leq$50\% complete at $\muv>-17.5$.
		}
		\label{fig:completenessfull}
	\end{figure}

	The completeness of our analysis sample is depicted in Figure \ref{fig:completenessfull} as a function of \muv{} (the measurement of which is described in Section \ref{sec:luv}). The parent sample is represented by the objects targeted by \textit{JWST}/NIRSpec within the JADES and CEERS surveys, including objects that did not have recovered spectroscopic redshifts within our reduction. The majority of the sample incompleteness is driven by low S/N \hb{} and a lack of any strong emission lines that prevent a spectroscopic redshift measurement. Our sample is $66$\% complete at $\muv<-19.5$, with the caveat that this completeness analysis does not include objects in which no photometric redshift was available (N=15) and objects without photometric coverage near $\lambda_{\rm rest}\sim1500$\AA{} (and thus, no estimate of \muv{}, N=15). A further discussion of how completeness affects our analysis can be found in Section \ref{sec:z_complete}.
	
\begin{figure}
	\centering
	\includegraphics[width=\columnwidth]{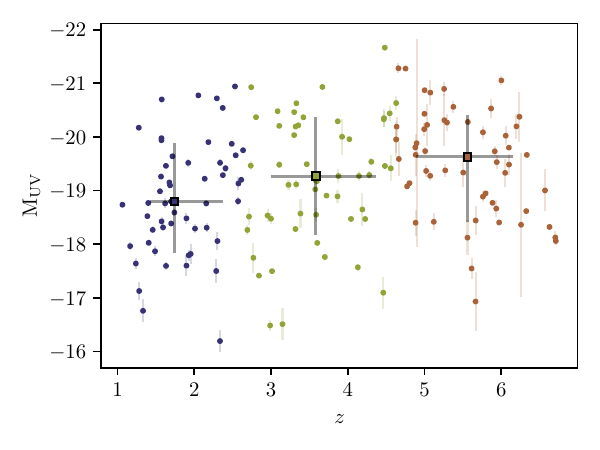}
	\caption{The redshift and \muv{} distribution of the combined CEERS/JADES NIRSpec sample. The sample is split into three redshift bins, with the objects in the lowest redshift bin (N=54) shown in dark blue, the objects in the middle bin (N=55) shown in green, and the objects at the highest redshifts (N=54) shown in brown. The median \muv{} and $z$ are displayed as larger squares, with errorbars on the medians representing the inner 68th percentile range of each subsample.
	}
	\label{fig:sample}
	
\end{figure}

	We display the redshift and UV absolute magnitude distribution of our full analysis sample of 163 galaxies in Figure \ref{fig:sample}. We split the sample into three equal bins of increasing redshift, and color code each object according to its bin membership. The lowest redshift bin features 54 objects at $1.06<z\leq2.64$ and $z_{\rm med}=1.74_{-0.32}^{+0.63}$, the second redshift bin features 55 objects at $2.64<z\leq4.63$ and $z_{\rm med}=3.59_{-0.59}^{+0.78}$, and the highest-redshift bin features 54 objects at $4.63<z\leq6.71$ and $z_{\rm med}=5.56_{-0.68}^{+0.59}$. The errorbars represent the 16th and 84th percentile of the subsample redshift distribution. \muv{} evolves mildly within our sample: $-18.80_{-1.10}^{+0.96}$, $-19.27_{-1.10}^{+1.09}$, and $-19.63_{-0.78}^{+1.22}$ are the median \muv{} for the three increasing redshift bins.

	\subsection{\xiion{}}
	
	The Lyman continuum production efficiency is defined as the ratio of the production rate of ionizing photons ($\dot{n}_{\rm ion}$) in units of s$^{-1}$ to the intrinsic UV continuum luminosity density ($L_{\rm UV,int}$) in units of erg\:s$^{-1}$\:Hz$^{-1}$:
	\begin{equation}
		\xiion = \dfrac{\dot{n}_{\rm ion}}{L_{\rm UV,int}} [\textrm{erg}\:\textrm{Hz}^{-1}]^{-1}
	\end{equation}
	In this work, $L_{\rm UV,int}$ is measured at 1500\AA{} (described in Section \ref{sec:luv}) and $\dot{\rm n}_{\rm ion}$ is calculated from the dust-corrected \ha{} luminosity ($L_{\textrm{H}\alpha\textrm{,int}}$, described in Section \ref{sec:int_ha}). Assuming no ionizing photons escape the galaxy ($\fesc=0$) and case B recombination, we can relate $\dot{\rm n}_{\rm ion}$ to $L_{\textrm{H}\alpha\textrm{,int}}$ via the relation of \citet{leithererSyntheticPropertiesStarburst1995}:
	\begin{equation}
		\dot{\rm n}_{\rm ion} [\textrm{s}^{-1}] = \dfrac{1}{1.36} \times 10^{12} L_{\textrm{H}\alpha\textrm{,int}} [\textrm{erg}\:\textrm{s}^{-1}]
	\end{equation}
	This equation assumes a temperature of $10^4$K and an electron density of 100\:cm$^{-3}$, but is not sensitive to the choice of stellar population model. A nonzero \fesc{} would boost the derived \xiion by a factor of 1/(1-\fesc{}). We discuss various \fesc{} assumptions in Section \ref{sec:reionization}, and present the majority of results within this work as \xiiono{} (assuming \fesc=0).
						
	\subsubsection{\luv{} and the stellar dust correction} \label{sec:luv}
	
	In order to constrain the intrinsic non-ionizing UV luminosity, we first estimated the observed \muv{} at 1500\AA{} using the best-fit SED models. For each object, we first converted the best-fit model to the rest frame. We then averaged the flux density within a 100\AA{} window around $\lambda_{\rm rest}=1500$\AA{} and converted this flux density to an absolute AB magnitude. In order to estimate the uncertainty on \muv{}, we averaged the errorbars on photometric datapoints neighboring rest-frame 1500\AA{} and applied them to the \luv{} measurement. This methodology ensured a consistent rest-wavelength location for each estimate of \muv{}, with errorbars motivated by nearby photometric constraints. As mentioned in Section \ref{sec:selection}, we require photometric coverage of $\lambda_{\rm rest}<2000$\AA{} and a well-constrained best-fit SED model for inclusion within our analysis sample.
	
	The \muv{} measurements were corrected for stellar dust attenuation using the \av{} best-fit parameter from the SED fitting. For each object, we assumed either a \citet{calzettiDustContentOpacity2000} or an SMC dust attenuation curve \citep{gordonQuantitativeComparisonSmall2003}, depending on the stellar mass and redshift of the object, in order to be consistent with that assumed by the SED fit. We applied the following correction to the observed \luv{} measurements:
	\begin{equation} \label{eqn:luvint}
		L_{\rm UV,int} = 10^{k_{\lambda} \times \textrm{A}_{\rm V} / (2.5 R_{\rm V})} L_{\rm UV,obs},
	\end{equation}
	where $R_{\rm V}=2.74$ for the SMC dust attenuation curve and $R_{\rm V}=4.05$ for \citet{calzettiDustContentOpacity2000}.
	
	\subsubsection{$L_{\textrm{H}\alpha\textrm{,int}}$ and the nebular dust correction} \label{sec:int_ha}
	
	To correct the observed \ha{} luminosity for nebular dust attenuation, we assumed a \citet{cardelliRelationshipInfraredOptical1989}  Galactic extinction curve and E(B-V) values derived from the \ha{}/\hb{} Balmer decrement. We assumed an intrinsic ratio of \ha{}/\hb{} of 2.79, corresponding to Case B recombination where $T_{\rm e}=15,000$K \citep{reddyEffectsStellarPopulation2022}. In cases where the \ha{}/\hb{} flux ratios were below 2.79, we assumed E(B-V$)=0$.
	
	\subsubsection{Uncertainty on \xiion{}}
	
	We employed Monte Carlo simulations in order to estimate the uncertainties on each \xiiono{} measurement. For each run, we perturbed the observed \luv{} according to its error, then chose a random $A_{V}$ according to the distribution of best-fit $A_{V}$ output by FAST during the SED fitting. We corrected this observed \luv{} into an intrinsic \luv{} according to Equation \ref{eqn:luvint}. We then generated an $\dot{n}_{\rm ion}$ value after perturbing both the \ha{} and \hb{} fluxes according to their error, and calculated a final \xiiono{} for each run. This procedure was repeated 10,000 times, and the errors on \xiiono{} were calculated from the 16th and 84th percentiles of the resulting distribution. We determined the final  \xiiono{} measurement as the median of the distribution.
	
	\subsection{Equivalent widths}
	
	We wish to compare our \xiion measurements to a suite of galaxy properties, including the equivalent width (EW) of the \oiiif{} nebular emission line, which has been known to be strong tracer of \xiion{} \citep[e.g.,][]{chevallardPhysicalPropertiesHionizingphoton2018,tangMMTMMIRSSpectroscopy2019}. In order to constrain the equivalent width of this emission line, we require line fluxes, as detailed in Section \ref{sec:lineflux} and \citet{clarkeStarFormingMainSequence2024}, and an estimate of the flux density of the continuum near the emission line. In the majority (150/168) of objects within our sample, the continuum was not significantly detected (average SNR < 3)  in the NIRSpec spectrum near \oiiif{} . Thus, we used the best-fit SED models, fit to the emission-line corrected SEDs, to estimate the continuum for equivalent width measurements. For the 18 objects with a well-detected rest-optical continuum, we measured the flux density in the spectral range $\lambda_{\rm rest}=5027$\AA$-5067$\AA, and found that the resulting EWs are consistent within $4\%$ on average to those measured with the best-fit SED. Considering the match between the spectral and SED-based EWs, we used flux densities from best-fit SED models for all objects in order to calculate equivalent widths for consistency.
	
	\subsection{Metallicities}
	
	A remarkable number of strong emission lines were recovered for these objects thanks to the depth and wavelength coverage of the \textit{JWST}/NIRSpec observations. In order to examine \xiiono{} as a function of metallicity, we utilized strong-line ratios that include \oiii{}, \oii{}, \ha{}, \hb{}, and \neiii{} in order to estimate metallicity using strong-line diagnostics calibrated with early JWST spectroscopy. We use the metallicity diagnostics developed in \citet{sandersDirectEbasedMetallicities2024}, which are presented as a function of the ratios of R23, O32, O3, O2, and Ne3O2. Metallicities are calculated according to the method presented in \citet{sandersMOSDEFSurveyEvolution2021}, in that they are determined using a $\chi^2$ minimization across all strong-line diagnostics that were available for each object simultaneously \citep[see equation 6 of][]{sandersMOSDEFSurveyEvolution2021}. We set upper and lower limits on the metallicity at $12+\textrm{log(O/H}=7.2$ and  $12+\textrm{log(O/H}=8.5$, according to the range of individual metallicities used to generate the diagnostics of \citet{sandersDirectEbasedMetallicities2024}. 
	
	We utilized Monte Carlo simulations to determine the errorbars on  $12+\textrm{log(O/H)}$, by perturbing each line flux 200 times according to its error, re-measuring each strong-line ratio, then re-determining $12+\textrm{log(O/H)}$. Errors on individual metallicities measurements are presented as the 16th and 84th percentiles of the distribution, with the final $12+\textrm{log(O/H)}$ measurement taken as the median. If the inner 68th percentile of the $12+\textrm{log(O/H)}$ distribution was either below 7.2 or above 8.5, these measurements were assumed to be lower or upper limits, respectively.
	
	\subsection{UV slope}
	
	The UV continuum slope ($\beta$) has been considered a proxy for \xiion{}, particularly before \textit{JWST} gave direct access to Balmer line measurements at $z\gtrsim2.7$ \citep[e.g., ][]{robertsonNewConstraintsCosmic2013,duncanPoweringReionizationAssessing2015}.We constrained $\beta$ for objects within our sample by performing a linear regression on the equation 
	\begin{equation}
		m_{\rm AB} = -2.5(\beta+2)\times \textrm{log}_{10}(\lambda) + C,
	\end{equation}
	where $C$ is treated as a normalization constant, and $m_{AB}$ includes all photometric data points between $\lambda_{\rm rest}=1250$\AA{}$-2600$\AA{}. We required at least two photometric data points in this wavelength range. We perturbed the photometry according to the uncertainties 10,000 times, and found the median, 16th percentile, and 84th percentile of the resulting distribution to calculate $\beta$ and its uncertainties.
	
	\section{\xiion{} in the combined CEERS/JADES NIRSpec sample} \label{sec:res}
	
\begin{figure*}[hbpt!] 
		\centering
		\includegraphics[width=\textwidth]{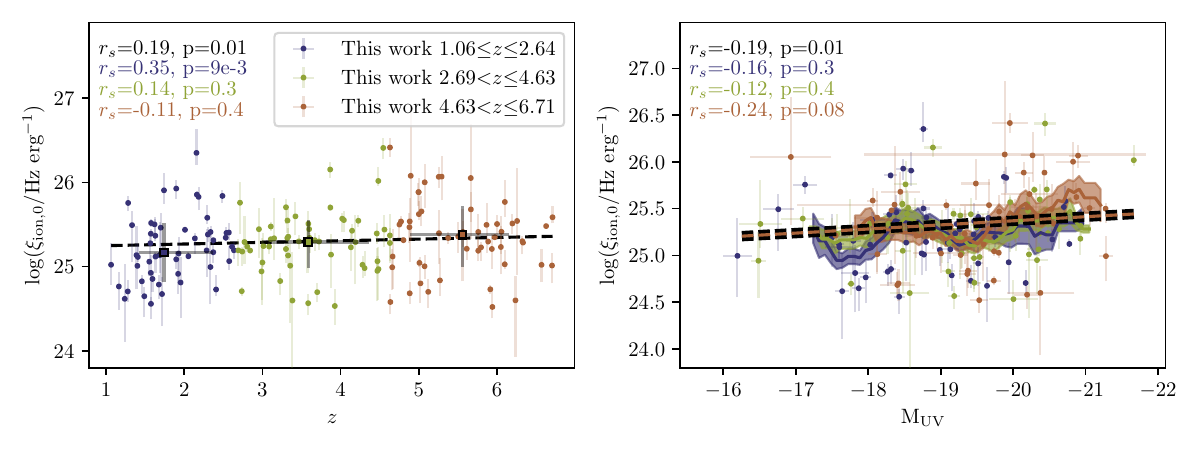}
		\caption{Trends between \xiiono{}, $z$, and \muv{} within the CEERS/JADES sample.
		\textbf{Left:} \xiiono{} as a function of redshift. Objects are color coded according to three bins of increasing redshift, with each bin containing approximately equal galaxies. Medians \xiiono{} of each redshift subsample are displayed as larger squares, with errorbars representing the 16th and 84th percentile of the \xiiono{} distribution. A multivariate fit to \xiiono{}, $z$, and \muv{} (Equation \ref{eqn:zmfit}) is displayed as a dashed, black line, and is evaluated at $\muv=-19.27$. Spearman correlation coefficients and $p$ values between \xiiono{} and $z$ are displayed in the upper left corner. From top to bottom, these coefficients are calculated from the full sample, objects in the lowest redshift bin, objects in the middle redshift bin, and objects in the highest redshift bin.
		\textbf{Right:} \xiiono{} as a function of \muv{}. Objects are color coded as in the left panel, and Spearman correlation coefficients and $p$ values are displayed in a similar manner. A rolling average of \xiiono{} is shown as a solid line, and is calculated and color coded for objects within each redshift sample. The window size for the rolling mean is 1/12 of the dynamic range in \muv{} for each subsample. The shaded region represents the standard error on the mean within each window. A multivariate fit to \xiiono{}, $z$, and \muv{} (Equation \ref{eqn:zmfit}) is displayed as a dashed, brown, bolded line, and is evaluated at $z=5.56$.
		}
		\label{fig:xiion_z_M_uv}
		
	\end{figure*}

	We measure a median $\lxiiono=25.29^{+0.29}_{-0.37}$ within the full sample, with the errorbars on the median representing the percentile range containing the inner 68\% of the data. We display the redshift evolution of \xiiono{} within our sample in the left panel of Figure \ref{fig:xiion_z_M_uv}. We calculate the Spearman correlation coefficient between \xiiono{} and $z$, and find $r_{s}=0.19$ with a $p$-value of 0.01, representing a statistically-significant evolution of increasing \xiiono{} with redshift. We additionally calculate the median \xiiono{} within the three bins of increasing redshift detailed in Figure \ref{fig:sample}. At the lowest redshifts, with $z_{\rm med}=1.74_{-0.32}^{+0.63}$, we recover a median $\lxiiono=25.17_{-0.35}^{+0.34}$. Within the bin of $z_{\rm med}=3.59_{-0.59}^{+0.78}$, we find $\lxiiono=25.29_{-0.31}^{+0.26}$, and for the objects in the highest-redshift bin with $z_{\rm med}=5.56_{-0.68}^{+0.59}$, we find $\lxiiono=25.38_{-0.38}^{+0.35}$. We additionally recalculate the Spearman correlation coefficient for objects within each redshift bin, and only find a statistically-significant correlation within the lowest redshift bin, with $r_s=0.19$ and $p=9e-3$. Therefore, we cannot confidently state that there is evidence of evolution of $\xiiono$ at $z\gtrsim5$.
	
	We additionally show the relationship between \xiiono{} and \muv{} within our sample in the right panel of Figure \ref{fig:xiion_z_M_uv}. Here, we color code each galaxy with respect to its membership in the aforementioned bins of redshift. Across the entire sample, we calculate the Spearman correlation coefficient between \xiiono{} and \muv{} of $r_s=-0.19$ with a $p$-value of 0.01. Accordingly, we find that UV bright galaxies are more efficient producers of ionizing radiation with moderate significance. Within each redshift bin, we display the moving average of \xiiono{} with respect to \muv{} as a solid line in the right panel of Figure \ref{fig:xiion_z_M_uv}. While both moving averages and $r_s$ are largely consistent between redshift subsamples, we do not find any statistically-significant correlations between \xiiono{} and \muv{} at fixed redshift according to the $p$-values of $r_s$. We also note that our samples is $46$\% complete at $\muv{}<-19.5$: we discuss this incompleteness in Section \ref{sec:z_complete}.
	
	Given the apparent relationships between \xiiono{}, $z$, and \muv{}, we develop a framework for predicting \xiiono{} for a galaxy at a given epoch and UV brightness. We perform a multivariate orthogonal distance regression on the full sample using the python package \textsc{scipy}, and find the relation:
	\begin{multline} \label{eqn:zmfit}
		\textrm{log}(\xi_{\rm ion,0}/\textrm{Hz erg}^{-1})=(0.02\pm0.01)\times z \\ +(-0.04\pm0.02)\times \textrm{M}_{\rm UV}+24.38\pm0.41.
	\end{multline}
	We display this relation in Figure \ref{fig:xiion_z_M_uv} projected into $z$ and \muv{} space. In the left panel, we evaluate Equation \ref{eqn:zmfit} at the full sample median $\muv=-19.27$ and display it as a dashed, black line, and in the right panel, we evaluate the relation at the median $z=5.56$ of the highest redshift bin and display it as a dashed, brown line. Given that we do not find evidence of evolution between \xiion{} and $z$ or \muv{} within smaller redshift subsamples, we caution extending this relationship beyond the dynamic range of our data. We additionally do not find significant correlations between \xiion{} and $z$ when splitting the full sample into three bins of increasing \muv{}, obfuscating which of the two parameters ($z$ or \muv{}) is more fundamental in determining the evolution of \xiiono{}.
	
\begin{figure*}
		\centering
		\includegraphics[width=\textwidth]{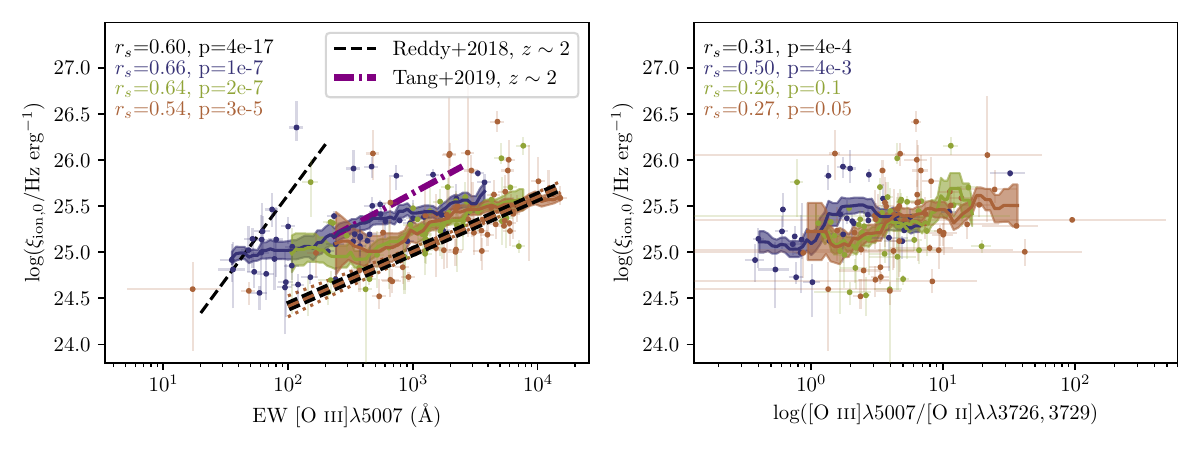}
		\caption{
			\xiiono{} as a function of the equivalent width of \oiiif{} (left) and O32 (right). Objects are color coded as in the redshift subsamples in Figure \ref{fig:xiion_z_M_uv}, and Spearman correlation coefficients and rolling means are displayed in a similar manner. A linear relation between \xiiono{} and \woiii{} fit with the objects in the highest redshift subsample (Equation \ref{eqn:oiii}) is displayed as a brown, dashed, bolded line in the left panel. The \xiiono{}-\woiii{} relationships from the MOSDEF survey \citep{reddyMOSDEFSurveySignificant2018} and from \citet{tangMMTMMIRSSpectroscopy2019} at $z\sim2$ are displayed as a dashed, black line and a dash-dot, purple line, respectively. A strong, significant correlation between both \xiion{} and \woiii{} and \xiion{} and O32 is found within our sample.
		}
		\label{fig:xiion_O3}
		
	\end{figure*}

	We find the strongest and most significant correlations between \xiion{} and properties related to the strength of [OIII] emission. In the left panel of Figure \ref{fig:xiion_O3}, we show \xiiono{} vs. \woiii{} within the CEERS/JADES NIRSpec sample, with galaxies again color-coded according to three redshift bins. We find $r_s=0.60$ across the full sample, with $p=4$e-17, representing a highly-monotonic correlation between \xiiono{} and \woiii{} with strong significance. There are also correlations with similar $r_s$ at high significance within each redshift sample. We display moving averages for each redshift bin as solid lines in Figure \ref{fig:xiion_O3}: these averages indicate that there is less-efficient production of ionizing photons at fixed \woiii{} when moving towards higher redshift, although higher-redshift objects have higher \woiii{} and \xiiono{} overall. We also find a highly-significant correlation between \xiion{} and O32, defined as the ratio of \oiiif{} and \oii{} line fluxes, shown in the right panel of Figure \ref{fig:xiion_O3}, with $r_s=0.31$ and $p=4$e-4. A correlation between \xiion{} and O32 can be explained by high ionization parameters, as well as an anti-correlation between O32 and metallicity \citep[e.g.]{jonesTemperaturebasedMetallicityMeasurements2015,bianDirectGasphaseMetallicity2018,curtiMassmetallicityFundamentalMetallicity2020,sandersDirectEbasedMetallicities2024}.

	\woiii{} has been demonstrated as a powerful diagnostic for \xiion{}, with relationships developed by observational studies both at $z\sim0$ \citep{chevallardPhysicalPropertiesHionizingphoton2018} and at $z\sim2$ \citep{tangMMTMMIRSSpectroscopy2019,reddyMOSDEFSurveySignificant2018}. Given an apparent evolution of the trend with redshift as shown in the left panel of Figure \ref{fig:xiion_O3}, we calculate a novel trend between \xiiono{} and \woiii{} calculated entirely from objects at $z\geq4.63$.  We perform an orthogonal distance regression using the python package \textsc{scipy} and find the relation
	\begin{multline} \label{eqn:oiii}
		\textrm{log}(\xi_{\rm ion,0}/\textrm{Hz erg}^{-1})=(0.59\pm0.09)\times \textrm{log}(\woiii) \\ +23.23\pm0.31.
	\end{multline}
	We display Equation \ref{eqn:oiii} in Figure \ref{fig:xiion_O3} as a brown, dashed, bolded line. The 95\% confidence region for this linear regression is small enough to fit within the bolded line. We additionally display \xiion{}-\woiii{} relationships found at $z\sim2$ within the MOSDEF survey \citep{reddyMOSDEFSurveySignificant2018} and a survey of extreme [O~\textsc{iii}] emitters \citep{tangMMTMMIRSSpectroscopy2019} in Figure \ref{fig:xiion_O3}: the redshift evolution of the \xiiono{}-\woiii{} relationship can be clearly seen when comparing Equation \ref{eqn:oiii} to these curves from the literature. 
	
\begin{figure*}
		\centering
		\includegraphics[width=\textwidth]{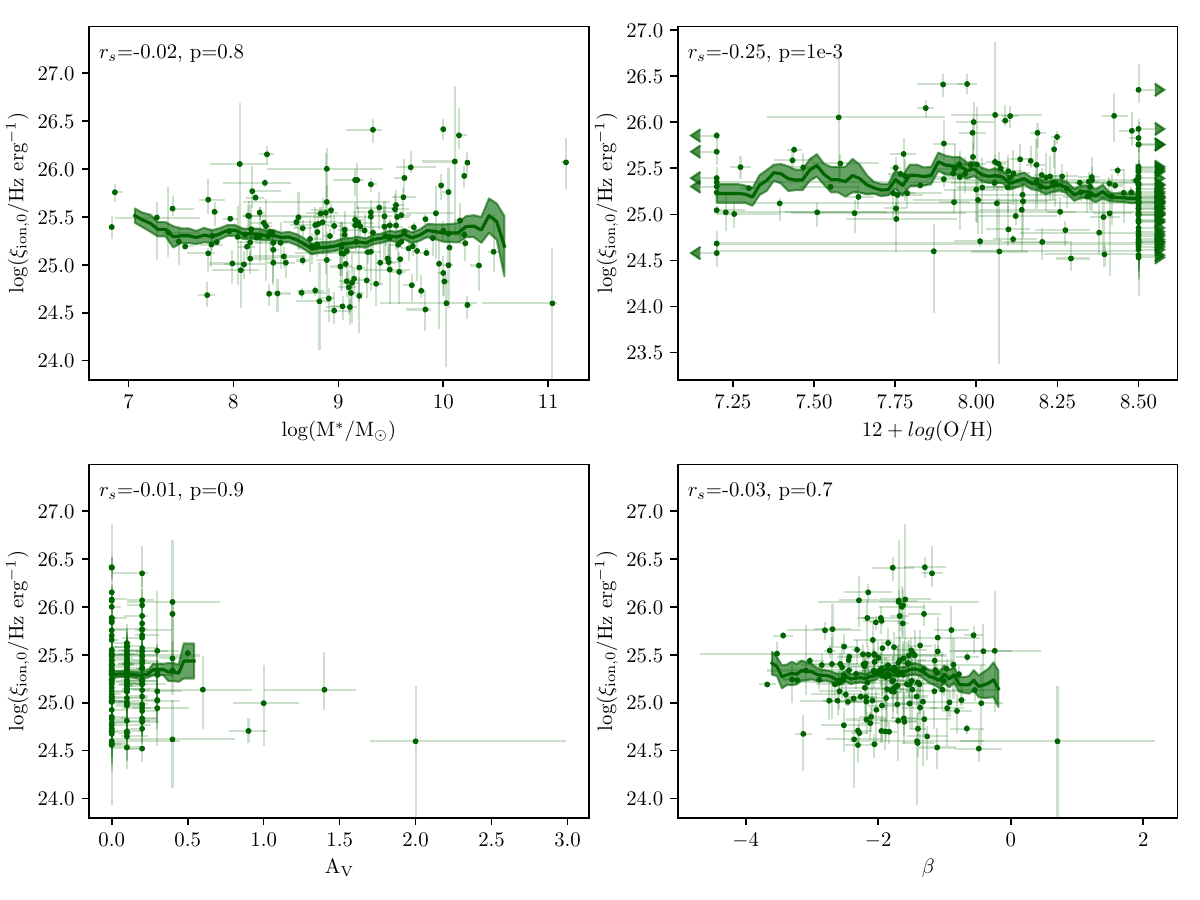}
		\caption{\xiiono{} as a function of stellar mass (\mstar{}), metallicity (\oh{}), dust attenuation (\av{}), and UV slope ($\beta$) for the full sample. Spearman correlation coefficients and corresponding $p$ values are displayed in the upper left of each panel. Rolling means are calculated in a similar manner to Figure \ref{fig:xiion_z_M_uv} and are displayed as solid, green lines, with shaded regions representing the standard error on the mean. We recover a significant anti-correlation between \xiiono{} and metallicity within our sample, and no significant correlation between \xiiono{} and \mstar{}, \av{} or $\beta$.
		}
		\label{fig:xiion_other}
	\end{figure*}

	We explore the relationships between \xiion{} and \mstar{}, \oh{}, \av{}, and $\beta$  in Figure \ref{fig:xiion_other}.  We find no significant correlation between \xiiono{} and \mstar{}, \av{} or $\beta$ according to the Spearman correlation test. We do find an $r_s=-0.25$ and a $p$-value of 1e-3 between \xiiono{} and \oh{}, indicating that objects with lower metallicities are more efficient producers of ionizing photons with a statistically-significant rejection of the null hypothesis (no correlation). Considering that $\beta$ may be simultaneously modulated by stellar ages \citep{leithererStarburst99SynthesisModels1999} and dust attenuation, and that there is no significant trend between \xiiono{} and \av{} within our sample, we also determine the correlation coefficient between \xiiono{} and $\beta$ for objects with $\av=0$ (N=62).  We find $r_s=-0.13$ with $p=0.3$, indicating no evidence of correlation between \xiiono{} and $\beta$ in a dust-free population. This lack of correlation is surprising considering the ubiquity of the trend found at $z\sim2$ \citep{shivaeiMOSDEFSurveyDirect2018}, however, a strong trend between \fesc{} and $\beta$ \citep{pahlConnectionEscapeIonizing2023,chisholmFarUltravioletContinuumSlope2022} would simultaneously induce a trend between \xiion{} and $\beta$. Here, we only present \xiiono{}, assuming $\fesc=0$ across the entire sample.

	\section{Discussion} \label{sec:discussion}
	
	\subsection{Comparison to literature}
	
	We show our combined CEERS/JADES NIRSpec \xiiono{} measurements as a function of redshift in the upper panel of Figure \ref{fig:xiion_lit}, alongside a selection of \textit{JWST}-based \xiion{} measurements from the literature. We also include the pre-\textit{JWST} \xiiono{} measurements from the MOSDEF survey at $z\sim2$ \citep{shivaeiMOSDEFSurveyDirect2018}, choosing to display results assuming an SMC dust extinction curve and $Z=0.2Z_{\odot}$ for consistency with the assumptions made for the majority of objects in this work.
	We recover an increasing trend between \xiiono{} and redshift between $z\sim2$ and $z\sim5$, consistent with the elevated \xiiono{} values found within the reionization era with \textit{JWST} \citep{mattheeEIGERIIFirst2023, endsleyStarformingIonizingProperties2023,prieto-lyonProductionIonizingPhotons2023,tangJWSTNIRSpecSpectroscopy2023,simmondsLowmassBurstyGalaxies2024,saxenaJADESProductionEscape2024}. We also show consistent mean \xiiono{} at the lowest redshifts of our sample as compared to the large-scale, ground-based spectroscopic survey of MOSDEF. We find no evidence of \xiiono{} increasing with redshift within the highest-$z$ bin of our sample, similar to the lack of correlation found within a sample of 17 faint Ly$\alpha$ emitters at $z>5.8$ within JADES, with \xiion{} similarly estimated with Balmer lines observed by NIRSpec \citep{saxenaJADESProductionEscape2024}, although this sample has a larger \xiion{} than that found within our $z_{\rm med}=5.56_{-0.68}^{+0.59}$ bin. \citet{simmondsLowmassBurstyGalaxies2024} recovered a strong trend between \xiion{} and redshift in the reionization era using a sample of 677 $z\sim4-9$ JADES galaxies that cover a larger dynamic range in $z$ than our highest-redshift bin, with \xiion{} estimated photometrically from SED fits using \textsc{prospector} \citep{johnsonStellarPopulationInference2021} and inferred from flux excesses in NIRCam photometric bands with coverage of \ha{} and \oiii{}.
	
	We find significantly larger scatter in \xiiono{} for individual objects at fixed redshift than results from literature. The results herein represent the first large-scale spectroscopic study of \xiiono{} combining JWST/NIRSpec data from multiple early surveys, with a simple selection function only considering objects with detectable \ha{} and \hb{} emission. Indeed, the \xiion{} measurements displayed from \citet{boyettExtremeEmissionLine2024} focus on galaxies with extreme \oiiif{} or \ha{} emission (EW > 750\AA{}), limiting the selection function as compared to this work. Those from \citet{saxenaJADESProductionEscape2024} represent objects with detectable Ly$\alpha$ emission: only five objects in our sample have $>3\sigma$ Ly$\alpha$ detections. 
	Both \citet{simmondsLowmassBurstyGalaxies2024} and \citet{prieto-lyonProductionIonizingPhotons2023} report similar scatter in \xiiono{} with respect to redshift, and represent efforts to constrain \xiiono{} directly from \textit{JWST}/NIRCam photometry, either from SED fitting or from inferred flux excesses of \ha{} and/or \oiii{}. These studies lack the Balmer decrement nebular dust attenuation corrections offered by \textit{JWST}/NIRSpec coverage of multiple Balmer lines. The SED modeling performed by \citet{endsleyStarformingIonizingProperties2023} appears to indicate that a bursty, star-forming phase is critical for driving the \xiion{} values of high-redshift galaxies. In the \citet{endsleyStarformingIonizingProperties2023} anlaysis, similar to \citet{simmondsLowmassBurstyGalaxies2024} and \citet{prieto-lyonProductionIonizingPhotons2023}, \xiion{} is constrained from SED fits to \textit{JWST}/NIRCam photometry. Considering that \ha{} is sensitive to short ($\sim3$ Myr) timescales of star-formation activity, it is unsurprising to see a large scatter in the \xiiono{}-$z$ relationship. The comparison of Balmer-based SFRs and UV-based SFRs in \citet{clarkeStarFormingMainSequence2024} indicates that bursty star-formation histories are prevalent in the early Universe.
	
	Given that a number of \xiion{} analyses have focused on objects in the JADES survey, we can directly compare our measurements of \xiiono{} and those from works with published, individual measurements in order to discern any difference in methodology. We find 16 objects that match those in \citet{simmondsLowmassBurstyGalaxies2024}, and find that their \xiion{} measurements are 0.13dex higher on average than ours. We find six overlapping objects when comparing to the \citet{saxenaJADESProductionEscape2024} sample, finding an average difference in \xiion{} of 0.11dex (such that our \xiiono{} measurements are lower, on average). This discrepancy appears to be driven by larger \ha{} flux measurements reported in \citet{saxenaJADESProductionEscape2024}, as well as brighter UV magnitudes measured directly from the NIRSpec PRISM spectra. The simultaneous offset of both measurements appears to indicate a difference in flux calibration between spectroscopic reduction procedures.
		
	\subsection{Faint galaxies are inefficient producers of ionizing photons}
	
\begin{figure*}
		\centering
		\includegraphics[width=\textwidth]{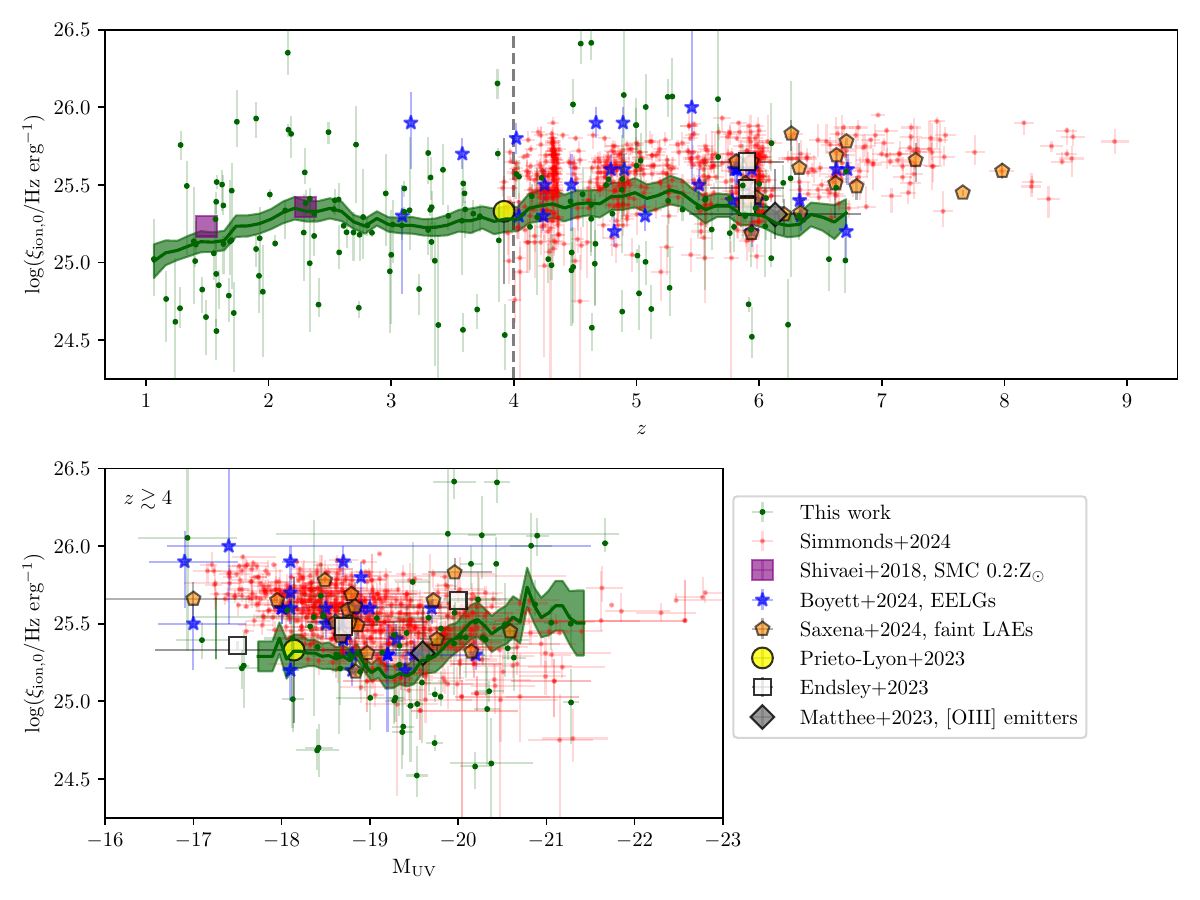}
		\caption{
			\textbf{Upper panel:} \xiiono{} as a function of redshift of the combined CEERS/JADES NIRSpec sample as compared to values in the literature. Rolling averages are calculated and displayed in a similar manner to that of Figure \ref{fig:xiion_z_M_uv}. \xiion{} values from the literature are both directly constrained from spectroscopy \citep{endsleyStarformingIonizingProperties2023,mattheeEIGERIIFirst2023,saxenaJADESProductionEscape2024,boyettExtremeEmissionLine2024} and estimated from photometry \citep{shivaeiMOSDEFSurveyDirect2018,prieto-lyonProductionIonizingPhotons2023,endsleyStarformingIonizingProperties2023,simmondsLowmassBurstyGalaxies2024}.
			\textbf{Lower panel:} \xiiono{} as a function of \muv{}, limited to \textit{JWST}-based high-redshift surveys. Only objects within our analysis sample with $z\geq4$ are displayed, in order to better compare to the redshift range of results from literature.
		}
		\label{fig:xiion_lit}
		
	\end{figure*}

	We restrict our sample to $z\geq4$ and display the corresponding \xiiono{} measurements with respect to \muv{} in the lower panel of Figure \ref{fig:xiion_lit}, alongside a variety of high redshift, \textit{JWST}-based \xiion{} measurements from the literature. We recover a mild trend of increasing \xiiono{} with increasing UV brightness within the full sample, although we find no significant trend within the highest-redshift population. The direction of this trend is consistent with that found in \citet{endsleyStarformingIonizingProperties2023}. The authors argue that the UV bright population in their work demonstrates larger instantaneous ($\sim3$Myr) star-formation rates as compared to those at $\sim50$Myr timescales, indicating that objects with higher UV luminosities have undergone a recent increase in SFR, while UV faint objects are experiencing recent declines. 
	
\begin{figure*}
		\centering
		\includegraphics[width=\textwidth]{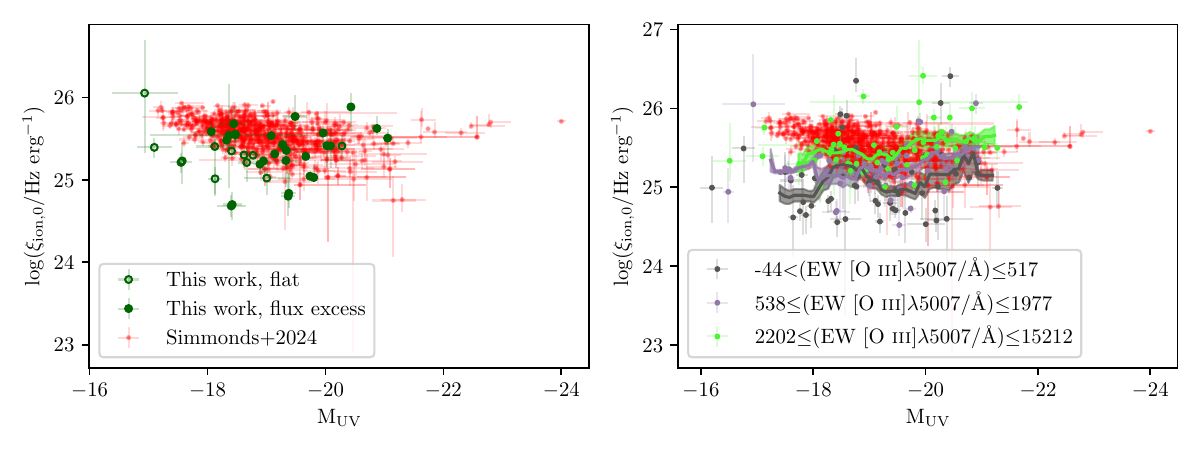}
		\caption{\xiiono{} vs. UV brightness for our analysis sample, with a variety of selection functions applied. The full sample analyzed by \citet{simmondsLowmassBurstyGalaxies2024} is shown as red points.
			\textbf{Left:} Objects within this work are displayed as solid, green points if they meet the photometric selection criteria of \citet{simmondsLowmassBurstyGalaxies2024}, which include flux excesses in filters containing \ha{} and \oiii{}. Objects are displayed as open, green points if no flux excesses are present in the requisite filters. Our analysis includes objects at fainter UV magnitudes and less-efficient \xiiono{} as compared to \citet{simmondsLowmassBurstyGalaxies2024}.
			\textbf{Right:} The full sample is color-coded according to \woiii{}. Objects are split into three equal bins of increasing \woiii{}, with lime green points representing objects in the highest-\woiii{} bin. Objects at the highest \woiii{} are most coincident with the \citet{simmondsLowmassBurstyGalaxies2024} \xiion{}-\muv{} distribution. 
		}
		\label{fig:xiion_S24}
		
		\end{figure*}

	An opposite relationship is recovered in \citet{simmondsLowmassBurstyGalaxies2024}, in that fainter objects were found to be more efficient producers of ionizing radiation. This conclusion is supported by the large \xiion{} values inferred for a sample of eight ultra-faint dwarf galaxies at $z\sim7$ \citep[$\xiion=25.80\pm0.14$,][]{atekMostPhotonsThat2024}. The sample of \citet{simmondsLowmassBurstyGalaxies2024}, unlike the similar JADES sample in \citet{endsleyStarformingIonizingProperties2023}, features a minimum flux difference between different wide and medium band JWST/NIRCam in order to probe significant \ha{} and \oiii{} fluxes within the photometry. We apply this redshift-dependent flux excess selection from \citet{simmondsLowmassBurstyGalaxies2024} to emission-line uncorrected photometry within our sample, and display objects that meet the requisite flux excess criteria ($>10$nJy) as solid points in the left panel of Figure \ref{fig:xiion_S24}. For objects with similar photometric measurements that do not meet the $10$nJy flux excess, i.e., objects that would be missed by the \citet{simmondsLowmassBurstyGalaxies2024} selection criteria, we display them as open points. Within our sample, we see that objects with too ``flat" photometry are on average fainter with lower \xiiono{}. The objects with flux excesses similar to the selection criteria of \citet{simmondsLowmassBurstyGalaxies2024} fall in similar \xiion{}-\muv{} parameter space to the \citet{simmondsLowmassBurstyGalaxies2024} objects, indicating that an observed anti-correlation between \xiion{} and UV brightness may be a selection effect. We additionally display our full sample in three increasing bins of \woiii{} in the right panel of Figure \ref{fig:xiion_S24}. We find that objects in the highest bin of \woiii{} are most coincident in \xiion{}-\muv{} space with objects in \citet{simmondsLowmassBurstyGalaxies2024}, indicating that a flux-excess-based selection criteria misses fainter objects with lower \woiii{}.

	\subsection{Implications for reionization: an adequate number of ionizing photons?} \label{sec:reionization}
	
	Given the apparent redshift evolution in \xiion{} found in early \textit{JWST} surveys, \citet{munozReionizationJWSTPhoton2024} developed a model of reionization that takes into account recent empirical constraints on the production and escape of ionizing photons within star-forming galaxies. The authors use UVLFs at $z<9$ from \citet{bouwensNewDeterminationsUV2021} and \textit{JWST}-based calibrations of \citet{donnanJWSTPRIMERNew2024} at $z\geq9$, \xiion{} trends with $z$ and \muv{} from \citet{simmondsLowmassBurstyGalaxies2024}, and either a fixed $\fesc=0.20$ or one tied directly to $\beta$, calibrated from $z\sim0$ LyC surveys \citep{chisholmFarUltravioletContinuumSlope2022}. Based on these assumptions, they find tension between the subsequent evolution of the neutral fraction and the CMB optical depth. In particular, they find that too many ionizing photons are produced, and reionization ends early ($z\sim8.5$, or $z\sim9.5$ assuming $\fesc=(1.3\pm0.6)\times10^{-4}\times10^{(-1.22\pm0.1)\beta}$). They invoke either a larger clumping factor (C=20), a lower $\fesc=0.03$, or a shallower faint-end slope of the UVLF as avenues to partially or completely explain this tension.
	
	The lower average \xiiono{} and trend of decreasing \xiiono{} with fainter \luv{} found in this work releases some of the tension identified by \citet{munozReionizationJWSTPhoton2024}. We recreate the reionization models of \citet{munozReionizationJWSTPhoton2024}, assuming an identical UVLF, a clumping factor of $C=3$, a case-B recombination coefficient evaluated at $T=2\times10^4$\:K, and a cut off in the UVLF at $\muv=-13$. We constrain the evolution of the ionizing emissivity based on different assumptions of \xiion(\muv,$z$) and \fesc(\muv{}), and compute the evolution of the neutral fraction as a function of redshift, while including effects from recombination.
	
\begin{figure*}
		\centering
		\includegraphics[width=0.84\textwidth]{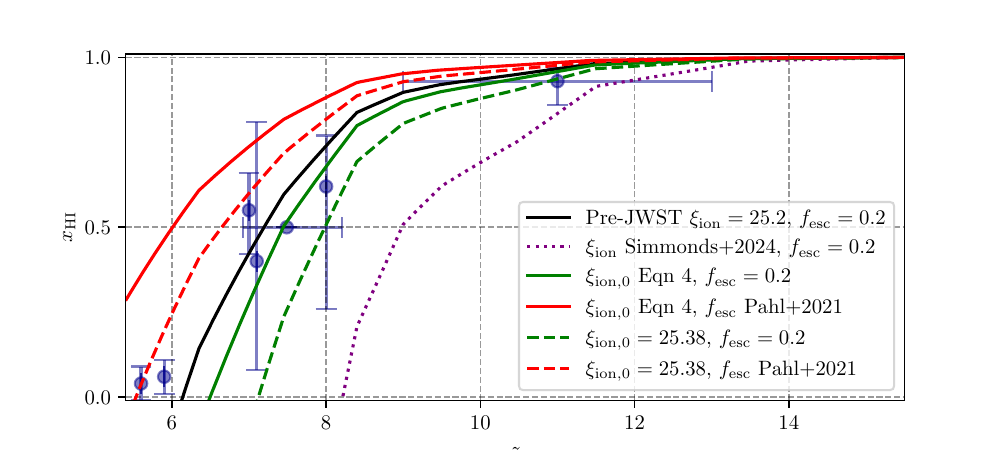}
		\caption{The evolution of the neutral fraction ($x_{\rm HI}$) according to recreations of the reionization models of \citet{munozReionizationJWSTPhoton2024}, with a variety of \xiion{} and \fesc{} assumptions. A solid, black line shows a model assuming $\lxiion=25.2$ and $\fesc=0.2$ (following \citet{robertsonCosmicReionizationEarly2015}). A purple, dotted line shows the same model assuming \xiion{} is a function of $z$ and \muv{} according to \citet{simmondsLowmassBurstyGalaxies2024}. A solid, green line shows a model assuming \xiiono{} evolves according to Equation \ref{eqn:zmfit} in this work, with the same assumed $\fesc=0.2$. A solid, red line shows the same with a fixed \xiiono{} corresponding to the median of our highest-redshift bin of objects ($z_{\rm med}=5.56_{-0.68}^{+0.59}$). Dashed lines show models assuming \xiiono{} motivated from this work and that \fesc{} is a two-valued function of \muv{} as found in the KLCS \citep{pahlUncontaminatedMeasurementEscaping2021}. Observational constraints on $x_{\rm HI}$ are shown as blue points \citep{mcgreerModelindependentEvidenceFavour2015,greigAreWeWitnessing2017,masonModelindependentConstraintsHydrogenionizing2019,whitlerImpactScatterGalaxy2020,nakaneLyaEmission132024}. 
		}
		\label{fig:xhi}
	\end{figure*}

	In Figure \ref{fig:xhi}, we show the evolution of the neutral fraction $x_{\rm HI}$ as a function of redshift based on recreations of the \citet{munozReionizationJWSTPhoton2024} reionization models. In black, we show the evolution with ``Pre-\textit{JWST}" values assumed by \citet{robertsonCosmicReionizationEarly2015}, namely that $\lxiion=25.2$ and $\fesc=20$\%. We also show a variety of observational constraints on $x_{\rm HI}$, including those from \citet{mcgreerModelindependentEvidenceFavour2015}, \citet{greigAreWeWitnessing2017}, \citet{masonModelindependentConstraintsHydrogenionizing2019}, \citet{whitlerImpactScatterGalaxy2020}, and \citet{nakaneLyaEmission132024}. The purple, dotted line shows the evolution of the neutral fraction following the \xiion{} presented \citet{simmondsLowmassBurstyGalaxies2024}, such that \xiion{} increases with $z$ and decreases with \muv{}. We cap \xiion{} at $z=9$ and $\muv=-16.5$ to avoid extrapolation. The LyC escape fraction, \fesc{} is set at $0.2$. Similar to \citet{munozReionizationJWSTPhoton2024}, we recover a reionization history that finishes significantly earlier compared to observational constraints when assuming  \citet{simmondsLowmassBurstyGalaxies2024} \xiion{} relations. In a solid, dark green line, we show the evolution of $x_{\rm HI}$ assuming \xiiono{} follows Equation \ref{eqn:zmfit}, such that \xiion{} increases with both redshift and \muv{}. To avoid extrapolation, we limit \xiiono{} at $z>6.71$, $\muv<-21.66$, and $\muv>-16.20$. We set \fesc{} at 0.20, which raises the \xiiono{} values presented here by 0.1dex. The resulting reionization history is comparable to that which assumes pre-\textit{JWST} canonical values for \xiion{}, and falls within the error on most observational constraints. Considering the lack of evolution found between \xiiono{}, $z$, and \muv{} in the highest redshift bin of our sample, we additionally display a dashed, green line which is a model assuming a flat \xiiono{} corresponding to the median of our highest-redshift bin, $\lxiiono=25.38$, and an $\fesc=0.2$.  This evolution is consistent with reionization ending at $z\gtrsim7$, which is disfavored by the majority of observational constraints on the ionization state of the IGM \citep[e.g.,][]{fanObservationalConstraintsCosmic2006,beckerMeanFreePath2021}.
	
	Adopting escape fractions that are tied to $\beta$ and calibrated from low ($z\sim0.3$) redshift LyC emitters and a $\beta-\muv{}$ relation drawn from \citet{zhaoDustHighRedshiftGalaxies2024}, the authors of \citet{munozReionizationJWSTPhoton2024} find that reionization ends significantly earlier than the curves presented in Figure \ref{fig:xhi}. The objects that make up the Low-redshift LyC survey (LzLCS) that were used to develop the $\fesc-\beta$ calibrations of \citet{chisholmFarUltravioletContinuumSlope2022} were selected based with either high O32, blue $\beta$, or high star-formation rate surface densities. The KLCS \citep{steidelKeckLymanContinuum2018,pahlUncontaminatedMeasurementEscaping2021} instead features 120 objects selected as Lyman break galaxies at $z\sim3$, with deep spectroscopy with coverage of the LyC observed for each object. Within the KLCS, an average $\fesc=0.06\pm0.01$ was recovered, with a strong correlation found between \fesc{} and \wlya{}. \fesc{} was thus found to be a roughly two-value function of \luv{}, such that objects with $\muv>-21$ had $\fesc=0.085\pm0.012$, and those with $\muv<-21$ had \fesc{} consistent with zero \citep{pahlUncontaminatedMeasurementEscaping2021}. We adopt these \fesc{} prescriptions as an alternative assumption within our reionization models, and display the resulting evolution of the neutral fraction as red curves in Figure \ref{fig:xhi}. These escape fraction assumptions represent avenues for a later reionization, particularly relevant for the fixed $\lxiiono{}$ assumption of 25.38, and are empirically motivated by observations of star-forming galaxies at some of the highest redshifts that \fesc{} measurements are possible \citep{vanzellaDetectionIonizingRadiation2012,steidelKeckLymanContinuum2018}.

	\subsection{Completeness in this work} \label{sec:z_complete}
	
	As discussed in Section \ref{sec:selection} and detailed in Figure \ref{fig:completenessfull}, our sample remains largely complete (66\%) only at the most luminous ($\muv<-19.5$) end of our sample. The majority of the incompleteness, however, is due to insufficient strength of either the Balmer H$\beta$ line, or a lack of any strong emission lines preventing a spectroscopic redshift measurement. Thus, the objects at the faint end of our sample likely represent an upper limit on \xiiono{} at fixed \muv{}, considering that objects with weaker \ha{}, \hb{}, and \oiii{} tend to be less efficient at producing ionizing photons. Additional objects that fall in the low-$\xiiono{}$ and faint \muv{} regime would strengthen the magnitude of our trend of decreasing \xiiono{} at fainter \muv{}, implying an even later reionization than the solid green line of Figure \ref{fig:xhi}. As demonstrated by this work, spectroscopy is critical to recover accurate \xiion{} estimates in the reionization era.
	Larger samples of lensed, high ($z\gtrsim6$) objects covered by deep, JWST/NIRSpec spectroscopy are necessary to bore out any potential evolution in the \xiiono{}-\muv{} relationship at the highest redshifts, with the potential to further reduce the current tension between the production of ionizing photons and the endpoint of reionization modeling.
		
		\section{Summary} \label{sec:summary}
		We use direct constraints on the intrinsic H$\alpha$ line luminosity, corrected for nebular dust attenuation via the Balmer decrement, and UV luminosity, constrained by best-fit SED models and corrected for stellar dust attenuation, in order to investigate \xiion{} in a sample of 163 objects drawn from both the CEERS and JADES surveys with deep \textit{JWST}/NIRSpec spectroscopy. Our main results are summarized below.
		
		\begin{enumerate}
			\item We find a median $\lxiiono=25.29^{+0.29}_{-0.37}$ within the full spectroscopic sample, with errorbars representing the 16th and 84th percentile of the distribution. Restricting our analysis to objects at the highest redshifts ($z>4.63$, $z_{\rm med}=5.56^{+0.59}_{-0.68}$, N=54), we find a median $\lxiiono{}=25.38^{+0.35}_{=0.38}$. Accordingly, we find a significant ($p=0.01$) correlation between \xiiono{} and $z$, with a Spearman correlation coefficient of $r_s=0.19$. An elevated \xiiono{} at higher redshifts is consistent with other \textit{JWST}-based analyses \citep[e.g.,][]{mattheeEIGERIIFirst2023,endsleyStarformingIonizingProperties2023,simmondsLowmassBurstyGalaxies2024}. We find no evidence for an evolution between \xiiono{} and $z$ within objects at $z>4.63$. We also find significantly-larger scatter in the \xiion{}-$z$ relationship than has been found in other works.
			\item We find a correlation between \xiiono{} and \muv{}, such that more UV bright objects have larger \xiiono{} ($r_s=0.19, p=0.01$). We perform a multivariate fit assuming \xiiono{} is both a function of $z$ and \muv{}, and find a slope between \xiiono{} and $z$ of $0.02\pm0.01$ and \muv{} of $-0.04\pm0.02$. While the direction of this \muv{} trend is consistent with some analyses of the JADES sample \citep{endsleyStarformingIonizingProperties2023}, \citet{simmondsLowmassBurstyGalaxies2024} finds an opposite trend, such that \xiiono{} increases towards fainter objects. We conclude that selection criteria based on photometric flux excesses may explain the difference in results between our analyses.
			\item We find that \xiiono{} is a strong function of both \woiii{} and O32, confirming results found both pre \textit{JWST} \citep[e.g.,][]{shivaeiMOSDEFSurveyDirect2018,chevallardPhysicalPropertiesHionizingphoton2018,tangMMTMMIRSSpectroscopy2019} and post \textit{JWST} \citep[e.g.,][]{boyettExtremeEmissionLine2024}. We develop a novel calibration between \xiiono{} and \woiii{}, fit to the objects at $z>4.63$ within our sample, and find that objects at higher redshifts have lower \xiiono{} at fixed \woiii{}. We explore \xiiono{} as a function of additional properties, and find no correlation between \xiiono{} and \mstar{}, \av{}, or UV slope. We do find a correlation between \xiiono{} and \oh{}, estimated from strong-line ratios.
			\item We examine the ionizing-photon budget crisis introduced by \citet{munozReionizationJWSTPhoton2024} in light of our novel constraints on \xiiono{}. Using \text{JWST} constraints on the high-redshift UVLF, constraints on \xiiono{} as motivated by our results, and a fiducial $\fesc=0.20$ we find that the tension between \textit{JWST}-era measurements of the production of ionizing photons and independent observational constraints on the neutral fraction are eased. We also explore escape fractions motivated directly from the $z\sim3$ Keck Lyman Continuum Spectroscopic survey in which \fesc{} is a two-valued function of \muv{}, and find avenues for reionization to end at $z\sim5-7$.
		\end{enumerate}
		According to our analysis, it's apparent that under assumed ionizing photon production efficiencies from our large-scale \textit{JWST}/NIRSpec analysis and $\fesc$ constrained at $z\sim3$, the ionizing-budget crisis induced by the larger \xiion{} found in early \textit{JWST} analyses is largely resolved. We highlight that our sample remains largely incomplete at the faint end of our analysis, but objects missed by our analysis are likely are inefficient producers of ionizing photons. Lensed, spectroscopic surveys of faint objects with \textit{JWST} will reveal a more complete picture of the \xiion{}-\muv{} relationship in the epoch of reionization.
		\linebreak
		\linebreak
		We acknowledge the CEERS and JADES teams for their effort to design, execute, and make public their observational surveys. 
		AJP would like to thank conversations with Gwen Rudie and Julian Mu\~noz, for discussions on science and reionization models, respectively.
		AJP was generously supported by a Carnegie Fellowship through the Carnegie Observatories while conducting this work.
		This work is based on observations made with the NASA/ESA/CSA James Webb Space Telescope. The data were obtained from the Mikulski Archive for Space Telescopes at the Space Telescope Science Institute, which is operated by the Association of Universities for Research in Astronomy, Inc., under NASA contract NAS5-03127 for JWST. Data were also obtained from the DAWN JWST Archive maintained by the Cosmic Dawn Center.
        The \textit{JWST} data used in this paper can be found in MAST: \dataset[10.17909/z7p0-8481]{doi:10.17909/z7p0-8481}, \dataset[10.17909/8tdj-8n28]{doi:10.17909/8tdj-8n28}.


	\end{document}